%% file: main_v3.tex
\documentclass[showpacs,superscriptaddress,aps,twocolumn,10pt,pre]{revtex4-2}

\usepackage{amsmath,amssymb}
\usepackage{graphicx}
\usepackage{dcolumn}
\usepackage{bm}
\usepackage{color,multirow}
\usepackage{subeqnarray}
\usepackage[colorlinks=true,%
            citecolor=blue,%
            linkcolor=blue,%
            urlcolor=blue]{hyperref}

\def\today{November 18, 2024; revised February 4, 2025}
\begin{document}
\newcommand{\blue}[1]{\textcolor{blue}{#1}}
\newcommand{\red}[1]{\textcolor{red}{#1}}
\newcommand{\green}[1]{\textcolor{green}{#1}}
\newcommand{\orange}[1]{\textcolor{orange}{#1}}

\title{Correction-to-scaling exponent for percolation and the 
       Fortuin--Kasteleyn Potts model in two dimensions}
\author{Yihao Xu}
   \email{yhxu@mail.ustc.edu.cn}
   \affiliation{Department of Modern Physics, 
                University of Science and Technology of China, Hefei, 
                Anhui 230026, China}	
\author{Tao Chen}
   \email{taochen@mail.ustc.edu.cn}
   \affiliation{Hefei National Research Center for Physical Sciences 
                at the Microscale, University of Science and Technology
                of China, Hefei 230026, China}
\author{Zongzheng Zhou}%
   \email{eric.zhou@monash.edu}
   \affiliation{School of Mathematics, Monash University, Clayton, 
                Victoria 3800, Australia}%
\author{Jes\'us Salas}
   \email{jsalas@math.uc3m.es}
   \affiliation{Universidad Carlos III de Madrid,
                Departamento de Matem\'aticas, 
                Avenida de la Universidad 30 (edificio Sabatini), 
                28911 Legan\'es (Madrid), Spain}
   \affiliation{Grupo de Teor\'{\i}as de Campos y F\'{\i}sica Estad\'{\i}stica,
                Instituto Gregorio Mill\'an, Universidad Carlos III de Madrid, 
                Unidad Asociada al Instituto de Estructura de la Materia, CSIC, 
                Serrano 123, 28006 Madrid, Spain}
\author{Youjin Deng}
   \email{yjdeng@ustc.edu.cn}
   \affiliation{Department of Modern Physics, University of Science 
                and Technology of China, Hefei, Anhui 230026, China}
   \affiliation{Hefei National Research Center for Physical Sciences 
                at the Microscale, University of Science and Technology 
                of China, Hefei 230026, China}
   \affiliation{Hefei National Laboratory, University of Science and 
                Technology of China, Hefei 230088, China}

\date{\today}

\begin{abstract}   
The number $n_s$ of clusters (per site) of size $s$, a central quantity 
in percolation theory, displays at criticality an algebraic scaling behavior 
of the form $n_s\simeq s^{-\tau}\, A\, (1+B s^{-\Omega})$. 
For the Fortuin--Kasteleyn representation of the $Q$-state Potts model in 
two dimensions, the Fisher exponent $\tau$ is known as a function of the 
real parameter $0\le Q\le4$, and, for bond percolation (the $Q\rightarrow 1$ 
limit), the correction-to-scaling exponent is derived as $\Omega=72/91$. 
We derive theoretically the exact formula for the correction-to-scaling  
exponent $\Omega=8/[(2g+1)(2g+3)]$ as a function of the Coulomb-gas
coupling strength $g$, which is related to $Q$ by $Q=2+2\cos(2 \pi g)$.
Using an efficient Monte Carlo cluster algorithm, we study the O($n$) loop 
model on the hexagonal lattice, which is in the same universality class as 
the $Q=n^2$ Potts model, and has significantly suppressed finite-size 
corrections and critical slowing-down. 
The predictions of the above formula include the exact value for 
percolation as a special case and agree well with the numerical estimates 
of $\Omega$ for both the critical and tricritical branches of the Potts model.  
\end{abstract}
	
\maketitle
	
	
%
%
\section{Introduction}  \label{sec:intro}

Bond percolation \cite{Stauffer_94,Bollobas_06} is a fundamental 
model in statistical mechanics, studying the formation and behavior of 
connected clusters in a lattice as a function of the probability 
$p \in [0,1]$ of each edge being occupied.  
In percolation, a key quantity is the size 
distribution $n_s(p)$, which gives the number of clusters (per volume) 
containing $s$ sites. Here, we are concerned with its behavior at the 
critical point $p_c$ \cite{Adler_82} 
\begin{equation}
  n_s(p_c) \;=\;  A\, s^{-\tau}\, \Big(1 + B\, s^{-\Omega} + \ldots \Big)\,.
  \label{def_ns_c}
\end{equation} 
where the dots stand for higher-order contributions. 
The Fisher exponent $\tau$ is universal and takes the value $\tau=187/91$ 
in two-dimensional (2D) bond percolation. The correction-to-scaling exponent 
$\Omega$ is also expected to be universal, and its value was precisely 
determined to be $72/91$ in a recent study \cite{Ziff_11} (see also 
Refs.~\cite{Ziff_99,Aharony_03}). 

Percolation can be defined on any $d$-dimensional lattice. However, in this
paper we focus on 2D lattices. More specifically, we consider models
defined on finite (undirected) graphs $G=(V,E)$ with vertex (or site) set $V$
and edge set $E$. In statistical mechanics, these graphs are usually taken as
finite subsets of a regular 2D lattice with some prescribed boundary 
conditions [e.g., toroidal in Monte Carlo (MC) simulations]. 
Therefore, the second-order phase transitions undergone by these
models will be governed in the infinite-volume limit by a certain conformal 
field theory (CFT) \cite{DiFrancesco_97}. 

Building on the percolation model, the random-cluster (RC) model 
\cite{Grimmett_06} introduces a more generalized framework. 
This model extends percolation theory by assigning a real weight 
$Q\ge 0$ to each connected cluster, including isolated sites. 
The partition function for the RC model on a finite graph $G=(V,E)$ 
is given by
\begin{equation}
   Z_{\text{RC}}(G;Q,p) \;=\; \sum_{F\subseteq E} p^{|F|}\, 
        (1-p)^{|E|-|F|}\, Q^{k(F)}\,, 
\label{def_Z_RC}
\end{equation}
where the sum is over all spanning subgraphs $(V,F)$ of $G$, $|F|$ represents 
the number of occupied edges, $|E|$ is the total number of edges, 
and $k(F)$ is the number of connected clusters. This model not only 
generalizes the percolation model (retrieved when $Q=1)$ but is also 
related to other significant statistical-mechanical models.  
	
The most famous connection is the $Q$-state Potts model 
\cite{Potts_52,Wu_82,Wu_82a,Wu_84,Baxter_85}, a widely studied model 
in statistical physics whose exact solution is still missing. 
In this model, each spin $\sigma_i$, located at a site $i\in V$ of 
the graph $G$, takes $Q\ge 2$ possible states, i.e., 
$\sigma_i \in \{1,2,\ldots,Q\}$. 
In this setup, $Q$ is an integer parameter. The Hamiltonian of this model 
is given by
\begin{equation}
   -\beta\, \mathcal{H}_{\text{Potts}}(\sigma) \;=\; 
    J \, \sum_{\{ij\} \in E} \delta_{\sigma_i,\sigma_j}\,,
\label{def_H_Potts}
\end{equation}
where $\sigma_i$ and $\sigma_j$ represent the states of neighboring sites 
$i$ and $j$, $\delta_{a,b}$ is the Kronecker delta function, $\beta$ is the
inverse temperature, and $J\in {\mathbb R}$ is the reduced nearest-neighbor 
coupling. Here we consider the ferromagnetic regime of this model $J > 0$.
Its partition function in this spin representation is given by
\begin{equation}
  Z_{\text{Potts}}(G;Q,J) \;=\; \sum\limits_{\{\sigma\}} 
    e^{-\beta\, \mathcal{H}_{\text{Potts}}(\sigma)}\,,
\label{def_Z_Potts}
\end{equation}
where the sum is over all possible spin configurations $\{\sigma\}$. 

This partition function \eqref{def_Z_Potts} can be rewritten in the 
so-called Fortuin--Kasteleyn (FK) \cite{Kasteleyn_69,Fortuin_72} 
representation as
\begin{equation}
   Z_{\text{Potts}}(G;Q,v) \;=\; \sum\limits_{F\subseteq E} v^{|F|}\, 
     Q^{k(F)}\,, 
\label{def_Z_FK}
\end{equation}
where the sum is over all spanning subgraphs $(V,F)$ of the graph 
$G$, $Q\ge 2$ is an integer, and the temperature-like variable 
$v=e^J-1$ belongs to the physical interval $v\in [0,\infty)$ in the
ferromagnetic regime.  
Due to the fact that \eqref{def_Z_FK} is a polynomial jointly in $Q$ and $v$,
we can promote these two variables from their physical ranges to
arbitrary real or even complex variables. In particular, if $Q,v>0$, the
model \eqref{def_Z_FK} has a probabilistic interpretation. 
The RC model \eqref{def_Z_RC} with $p=v/(1+v)$ 
directly maps onto the FK representation of the Potts model \eqref{def_Z_FK},
modulo some uninteresting prefactors.

The Potts model \eqref{def_H_Potts}/\eqref{def_Z_Potts} can be generalized
to include vacancies: these are basically integer variables 
$\tau_i \in \{0,1\}$ 
living on the sites of the graph, and such that $\tau_i=0$ (resp.\ $\tau_i=1$)
means that the site $i\in V$ is empty (resp.\ occupied). 
A simple Hamiltonian of this kind is given by 
\cite{Nienhuis_79,Nienhuis_80,Wu_82} 
\begin{equation}
-\beta\, \mathcal{H}_{\text{dPotts}}(\sigma,\tau) \;=\; 
    \sum_{\{i,j\}\in E} \tau_i \tau_j (K + J \delta_{\sigma_i,\sigma_j}) 
   - \Delta \sum_{i\in V} \tau_i \,, 
\label{def_H_diluted_Potts}
\end{equation}
where $\Delta$ is the chemical potential governing the concentration of 
vacancies. The corresponding partition function is
\begin{equation}
  Z_{\text{dPotts}}(G;Q,J,K,\Delta) \;=\; \sum\limits_{\{\sigma,\tau\}} 
    e^{-\beta\, \mathcal{H}_{\text{dPotts}}(\sigma,\tau)}\,,
\label{def_Z_diluted_Potts}
\end{equation}
where the sum is over all possible spin $\{\sigma\}$ and vacancy $\{\tau\}$ 
configurations. Other similar Hamiltonians have been considered in the 
literature \cite{Murata_79,Nienhuis_82a,Janke_04,Deng_05,Qian_05}.  
The diluted Potts model \eqref{def_H_diluted_Potts} appears naturally when 
performing a real-space renormalization-group (RG) transformation on the 
``pure''  Potts model \cite{Nienhuis_79,Nienhuis_80}.
The number of states $Q$ does not renormalize; it stays constant
along any RG trajectory. On the critical surface for $Q\in [0,4]$, 
there is a line of (attractive) critical fixed points (in the 
same universality class as the original Potts model), and there is another 
line of (repulsive) tricritical fixed points (belonging to a distinct 
universality 
class as their critical counterparts). Both lines meet at $Q=4$. For $Q> 4$, 
the system renormalizes to a discontinuity fixed point (at zero temperature),
as expected \cite{Nienhuis_75,Klein_76,Fisher_78}. 

Many critical exponents for the original and diluted $Q$-state Potts models
for $Q\in [0,4]$ are well known. One way of obtaining such exponents is by
relating both models to a Coulomb gas (CG)  \cite{Nienhuis_84,Nienhuis_87} with
a certain coupling constant $g$ (whose definition is not universal in the 
literature). In this paper, the parameters $Q$ and $g$ are related via 
\begin{equation}
  \sqrt{Q} \;=\; - 2\, \cos (\pi\, g)\,, \quad g \in \left( 0,2 
                                               \right]\,. 
\label{def_Q_vs_g}   
\end{equation} 
Both the critical and tricritical lines are covered by this parametrization:
the interval $g \in (0,1]$ (resp.\ $g\in [1,2]$) corresponds to the 
critical (resp.\ tricritical) line. Indeed, at $Q=4$ (or $g=1$) both lines
merge. 
For $g \in (0,1/2) \cup (3/2, 2]$, the value of $\sqrt{Q}$ is 
negative, but this is no an issue for the FK Potts model, 
as the partition function \eqref{def_Z_FK} depends on $Q=4\cos^2(\pi g) \ge 0$. 
For the O($n$) loop model (to be discussed later), this means that the 
statistical weight of each loop $n=\sqrt{Q}$ is negative and thus unphysical. 
Nevertheless, it is noted that the O($n$) loop model with negative weights 
can be well explored in the framework of CFT. 

The leading and subleading thermal exponents $y_{t1}$ and 
$y_{t2}$ relate to $g$ as (see e.g., Ref.~\cite{Nienhuis_87} and 
references therein) 
\begin{subeqnarray}
\slabel{def_y_t1}  
y_{t1} &=& \frac{3\, (2g-1)}{2\, g} \,, \\
y_{t2} &=& \frac{4\, (g-1)}{g} \,,  
\slabel{def_y_t2}  
\label{def_y_t1_and_t2}  
\end{subeqnarray} 
and the corresponding magnetic exponents $y_{h1}$ and $y_{h2}$ are
\begin{subeqnarray}
\slabel{def_y_h1}
y_{h1} &=& \frac{(2g+1)\, (2g+3)}{8\, g}\,, \\ 
y_{h2} &=& \frac{(2g-1)\, (2g+5)}{8\, g} \,.
\slabel{def_y_h2}
\label{def_y_h1_and_h2}
\end{subeqnarray}
The subleading thermal exponent \eqref{def_y_t2} corresponds to the so-called
dilution operator. This one is relevant for the tricritical Potts model and 
irrelevant for the critical Potts model.
At the critical 4-state Potts model, corresponding to $g=1$, that operator
is marginal with $y_{t2}=0$. More precisely, the dilution operator 
is in this case marginally irrelevant; this is the origin of multiplicative 
\cite{Nauenberg_80,Cardy_80} and additive \cite{Salas_97} logarithmic 
corrections. 
It is worth remarking that the dilution operator is responsible for
the aforementioned phase diagram for the dilute Potts model. This operator
is relevant (resp.\/ irrelevant) at the tricritical (resp.\/ critical) line, 
so, this line becomes repulsive (resp.\/ attractive). 

There are also exponents that are linked to the geometric properties of the 
FK clusters, like their fractal dimension $d_f$ \cite{Nienhuis_87}, 
which is equal to the leading magnetic exponent $y_{h1}$ \eqref{def_y_h1},  
\begin{equation}
  d_f \;=\; y_{h1} \;=\; \frac{(2g+1)(2g+3)}{8g} \,,
\label{def_df}
\end{equation}
and the so-called hull exponent \cite{Saleur_87} 
\begin{equation}
  d_H \;=\; \frac{1 + 2g}{2\, g} \,. 
\label{def_dH}
\end{equation}
Note that the above exponents, as well as many others, are rational
functions of $g$; but this is not true in general: the backbone exponent
for bond percolation $d_B$ has been proven to be a transcendental number 
\cite{Nolin_23}. 

We expect that the FK clusters in the critical Potts model, 
at least for $Q\in[1,4]$,
and the clusters in the percolation model corresponding to $g=2/3$
exhibit the same behavior of the size distribution
$n_s$ [cf. Eq.~\eqref{def_ns_c}], with a Fisher exponent $\tau$ given 
by the hyperscaling relation, and depending on $g$
\begin{equation}
\tau  \;=\; 1 + \frac{d}{d_f} \;\;\stackrel{d=2}{=}\;\;   
\frac{3+24g + 4g^2}{(1+2g)(3+2g)} \,,  
\label{def_tau}
\end{equation}
where $d=2$ is the dimensionality of the lattice. 

Our goal is to determine the correction-to-scaling exponent 
$\Omega(g)$ [cf. Eq.~\eqref{def_ns_c}] of the size distribution 
of the FK clusters for both the critical and tricritical Potts models. 
In Sec.~\ref{sec:Omega} we will show a theoretical argument leading to the 
conjecture
\begin{equation}
\Omega \;=\; \frac{1}{g\, d_f} \;=\; \frac{8}{(2g+1)\, (2g+3)} \,.
\label{conj_Omega}
\end{equation} 

In a second step, we want to check this result by using MC  
simulations \cite{Landau_14}. In this part, we will consider only integer 
values of $Q$. It is well known that MC algorithms suffer from 
critical slowing-down (CSD): roughly speaking, close to a critical point, the 
autocorrelation times diverge like $\xi^z$, where $\xi$ is the correlation 
length and $z$ is the dynamic critical exponent \cite{Sokal_97}. The 
mere existence of CSD limits the accuracy of the MC results.  
For local algorithms, CSD is severe: $z\approx 2$. 
A more efficient algorithm is the Swendsen--Wang (SW) cluster algorithm 
\cite{Swendsen_87} for the $Q$-state Potts ferromagnet with integer $Q\ge 2$. 
It radically reduces the value of $z$ but it cannot completely eliminate 
CSD whenever the specific heat diverges, due to the Li--Sokal bound
\cite{Li_89}
\begin{equation}
z \;\gtrsim\; \frac{\alpha}{\nu} \;=\; 2\, y_{t1}-2 \;=\; \frac{4 g - 3}{g}\,.
\label{def_sokal_li}
\end{equation} 
Thus, we have $z>0$ whenever $Q>2$ ($g>3/4$) for the Potts model.
Other cluster \cite{Chayes_98,Deng_07a} and worm-type
\cite{Prokofev_01,Deng_07b} algorithms for the Potts model also satisfy 
some kind of Li--Sokal bound; therefore, CSD is always present.

In any MC study, one usually needs to extrapolate from a series of 
numerical measurements of a physical quantity $Q_c(L)$ (performed for 
simplicity at the critical temperature) on 2D systems of linear size
$L\in \{L_1,L_2,\cdots,L_k\}$, to the behavior of $Q_c$ in the thermodynamic 
limit $L\to\infty$. This is achieved by performing fits to the ansatz
\begin{multline}
Q_c(L) \;=\; Q_{c,\circ} + 
             L^{p_Q}\, \left[a_0 + a_1\, L^{-\omega_1}   \right. \\ 
                       \left.    + a_2\, L^{-\omega_2} + \cdots 
\right]\,, 
\label{def_Ansatz} 
\end{multline}   
where $0< \omega_1 < \omega_2 < \cdots$ and the dots stand for 
higher-order corrections. It is clear that the larger the correction-to-scaling
exponents $\omega_i$, the more precise estimates of the three most relevant 
parameters $\{Q_\circ,p_Q, a_0\}$ will be obtained.  
These correction-to-scaling exponents not only come from the full set of 
subleading exponents [e.g., $y_{t2}$ \eqref{def_y_t2} or $y_{h2}$]. They   
are also generated by the nonlinear relations between the standard 
thermodynamic parameters (e.g., temperature, magnetic field, etc) and the 
RG nonlinear scaling fields (see \cite{Salas_00}, and references therein). 
These ``analytic'' corrections provide integer correction-to-scaling exponents. 
In addition, we have a particular model which merits special 
attention. The critical 4-state Potts model has $y_{t2}=0$, and this fact 
implies the existence of the aforementioned logarithmic corrections, which are 
very difficult to deal with. 

Therefore, instead of simulating the original \eqref{def_H_Potts} or the 
diluted \eqref{def_H_diluted_Potts} $Q$-state
Potts models to check conjecture \eqref{conj_Omega}, we simulate 
yet another model. The O$(n)$ loop model on a finite 2D hexagonal lattice $G$
\cite{Nienhuis_82,Batchelor_89,Peled_19,Duminil-Copin_21} 
is given by the partition function 
\begin{equation}
   Z_{\text{loop}}(G;x,n) \;=\; \sum_{\{\ell\}} x^{E(\ell)} \, 
   n^{N(\ell)}\,, 
\label{def_Z_loop}
\end{equation}
where the sum is over all possible nonintersecting loop configurations $\ell$
on $G$, $N(\ell)$ is the number of loops in the configuration, 
and $E(\ell)$ is the total length of all the loops. 
The parameter $n$ represents 
the weight associated to each loop, and $x$ is a fugacity that controls 
the weight of the loop's length. We assume here that $n,x$ are real 
positive parameters, so that \eqref{def_Z_loop} has a probabilistic 
interpretation. 

\begin{figure}[tbh]
\centering
\includegraphics[scale=0.65]{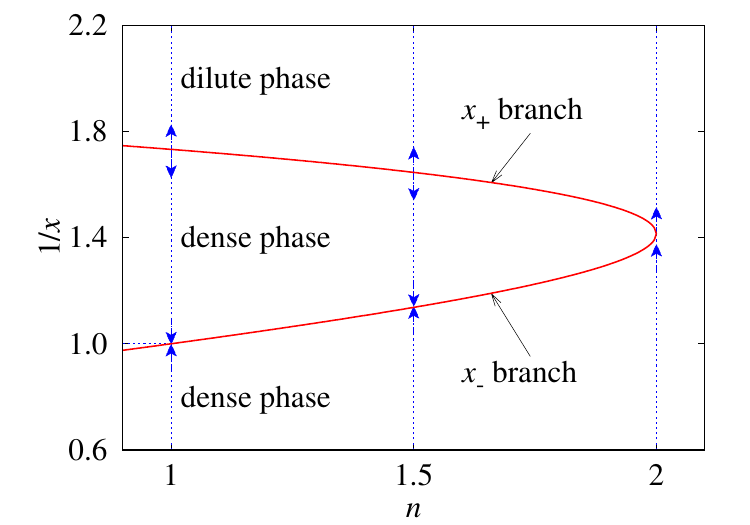}%
\caption{\label{fig:1} Phase diagram of the O($n$) loop model on the 
hexagonal lattice for $n \in [1,2]$. 
The system is in the dilute phase if the bond weight  
satisfies $x > x_{+}$, and in the dense phase if $x < x_{+}$. Vertical 
blue arrows sketch the directions of the RG flows. In 
terms of universality, the $x_{+}$ (resp.\ $x_{-}$) branch corresponds to 
the tricritical (resp.\ critical) Potts model.}
\end{figure}

We see in Sec.~\ref{sec:model} that the free energy of the O$(n)$ loop
model has been solved along two curves \cite{Nienhuis_82,Baxter_86,Baxter_87} 
\begin{equation}
   x_\pm \;=\; \frac{1}{\sqrt{2 \pm \sqrt{2-n}}} \,.
\label{def_xplusminus}
\end{equation}
These two curves are depicted in Fig.\ref{fig:1}. 
The curve $x_+(n)$ corresponds to the critical line of the O$(n)$ loop model
separating the diluted phase from the dense phase. The RG analysis of this 
model shows that the RG trajectories corresponds to constant values of $n$.  
Therefore, for a fixed value of $n\in [0,2]$, the interval 
$x\in (0,x_+(n))$ corresponds to the diluted phase, and $x \in (x_+(n),\infty)$,
to the dense (or densely packed) phase. 

Furthermore, the critical O$(n)$ loop model has a CG representation 
\cite{Nienhuis_84,Nienhuis_87} with coupling strength $g$, where
\begin{equation}
  n \;=\; - 2\, \cos (\pi\, g)\,, \qquad g \;\in\; [1,2]\,. 
\label{def_n_vs_g}   
\end{equation} 
Therefore, for a fixed value of $n\in [0,2]$, the point $(1/x,n)=(1/x_+(n),n)$
belongs the same universality class as the tricritical Potts model with
$Q=n^2$ states [cf. \eqref{def_Q_vs_g}/\eqref{def_n_vs_g}].

The leading thermal exponent for this model is given by \cite{Nienhuis_82}
\begin{equation}
y_{t1}^{\text{loop}} \;=\; y_{t2} \;=\; \frac{4\, (g-1)}{g} \,, 
\label{def_y_t1_On}
\end{equation}
where $y_{t2}$ is given by \eqref{def_y_t2}. Thus, the leading thermal 
eigenvalue for the O$(n)$ model is the first subleading thermal exponent for
the Potts model. 
The leading magnetic exponent is given by \cite{Nienhuis_82,Nienhuis_87}
\begin{equation}
y_{h1}^{\text{loop}} \;=\; \frac{(2+g)\, (2+3g)}{8\, g}\,.
\label{def_y_h1_On}
\end{equation}
Note that this expression is equal to $y_{h1}(1/g)$ [cf. \eqref{def_y_h1}].

Given a fixed value of $n \in [0,2]$, every point in the interval 
$1/x \in (0,1/x_+(n))$ renormalizes towards a fixed point that belongs to
the universality class of the critical Potts model with $Q=n^2$ states.  
In fact, $x_-$ [cf., Eq.~\eqref{def_xplusminus}] is the analytic extension 
of $x_+$ when $g\in (0,1]$. 
Thus, formulas 
\eqref{def_y_t1_On}/\eqref{def_y_h1_On} are also valid when 
$g\in (0,1]$. 

%
%

Notice that on both curves $x_\pm$, the leading eigenvalue is the subleading
eigenvalue $y_{t2}$ of the Potts model; thus it is smaller than the Potts 
leading one $y_{t1}$. In addition, on $x_-(n)$, $y_{t2} < 0$; i.e., it is 
irrelevant. Therefore, the correction-to-scaling corrections are smaller than
in the Potts model. Furthermore, the Li--Sokal bound becomes
\begin{equation}
z \;\gtrsim\; \frac{\alpha}{\nu} \;=\; 2\, y_{t1}-2 \;=\; 
              \frac{2\,(3 g - 4)}{g}\,.
\label{def_sokal_li_On}
\end{equation}
This implies that $\alpha/\nu \le -2$ on $x_-(n)$, and
$\alpha/\nu \le0$ on $x_+(n)$ for $n\ge 1$. Thus, there is no obstacle for
no CSD in this model. In this sense, highly efficient cluster algorithms for 
the O$(n)$ loop model with $n\ge 1$ have been proposed in the literature 
\cite{Deng_07,Ding_07,Liu_11,Fang_22}.

We may expect that the O$(n)$ loop model and the Potts model with $Q=n^2$
states that have the same CG coupling $g$ do belong to the same
universality class.  
The main assumption in this work is that the FK clusters of the critical and
tricritical Potts models have the \emph{same} critical behavior as the 
domains enclosed by the O$(n)$ loops.  
This hypothesis has been supported by analytic results for the fractal 
dimension $d_f$ or the one associated to the breakdown of cubic symmetry
$d_\text{CM}$ \cite{Nienhuis_87}. In \cite{Fang_22} the backbone and 
shortest-path exponents for the O$(n)$ loop model are shown to coincide 
in both models. Notice that they also conclude the absence of multiplicative
and additive logarithmic corrections for $Q=4$. Therefore, performing the MC
simulation on the O$(n)$ loop model has several advantages with respect to
standard simulations on the Potts model.  

Let us finish by comparing our work against 
Refs.~\cite{Aharony_03,Asikainen_03}. They consider corrections
to scaling to several quantities, e.g., the total mass of a cluster, as
a function of its radius of gyration $R$. They analyze several sources for
those corrections; CG arguments give the exponent $\theta'=1/g$. Indeed, if we
express \eqref{def_ns_c} in terms of the length scale $R \sim s^{1/d_f}$ [cf. 
\eqref{def_df}], we obtain that the correction term in \eqref{def_ns_c} is 
$(1 + b R^{-\omega} + \ldots)$ with $\omega=\Omega\, d_f = 1/g$. 

The remainder of this paper is organized as follows: Section~II
describes in detail the O$(n)$ model and its critical behavior.  
In Sec.~III, we show the theoretical argument leading to
our conjecture for $\Omega$ [cf. Eq.~\eqref{conj_Omega}]. 
Monte Carlo simulations and measurements are discussed in Sec.~IV, and 
in Sec.~IV, we present the numerical results. 
A brief discussion is given in Sec.~V.
	
%
%
\begin{figure}[tbh]
\centering
\includegraphics[scale=1]{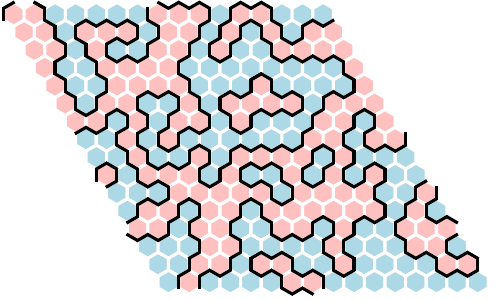}
\caption{\label{fig:2} Illustration of the correspondence between the 
O($n$) loop model on a hexagonal lattice with periodic boundary conditions 
and the generalized Ising model on the dual triangular lattice. 
The blue and red hexagons represent different Ising spins (up or down),
while the black lines depict the domain walls between the generalized 
Ising spins, which correspond to the configurations of the O($n$) loop model.
}
\end{figure}
 
%
%
\section{\texorpdfstring{The O$\bm{(n)}$ model}{The O(n)}}
\label{sec:model}

Let us consider a graph $G$ which is equal to a finite subset of the hexagonal
lattice with some boundary conditions. A loop in $G$ is a subgraph of $G$
which is isomorphic to a cycle. A loop configuration $\ell$ is a spanning 
subgraph of $G$ such as every vertex (or site) has even degree. 
Therefore, a loop configuration $\ell$ may contain isolated vertices. 
If we denote by $L(G)$ such a configuration space, the  
partition function of the O$(n)$ loop model 
\cite{Nienhuis_82,Batchelor_89,Peled_19,Duminil-Copin_21}
on $G$ is given by
\begin{equation}
   Z_{\text{loop}}(G;x,n) \;=\; \sum_{\ell \in L(G)} x^{E(\ell)} \, 
   n^{N(\ell)}
\label{def_Z_loop_Bis}
\end{equation}
[cf. \eqref{def_Z_loop}].
We start with real non-negative parameters
$x$ and $n$, so \eqref{def_Z_loop_Bis} has a probabilistic interpretation. As
\eqref{def_Z_loop_Bis} is a polynomial jointly in $x,n$, we can promote 
these variables to arbitrary real or complex variables. 
An O$(n)$ loop configuration on a hexagonal lattice with
periodic boundary conditions is depicted in Fig.~\ref{fig:2}. 

For $n=0$, one recovers formally the self-avoiding random walk model. 
For $n=1$, the O$(n)$ loop model is equivalent to the usual Ising model, 
and for $n=2$, it corresponds to the XY model. 

The O$(n)$ loop model has been solved by Nienhuis 
\cite{Nienhuis_82,Nienhuis_84} and 
Baxter \cite{Baxter_86,Baxter_87} on the curves \eqref{def_xplusminus}. 
In particular, there is a phase transition at $x_c(n) = x_+(n)$:
\begin{equation}
   x_c \;=\; x_+ \;=\; \frac{1}{\sqrt{2 + \sqrt{2-n}}} \,, \qquad 
         -2 \;\le \; n \;\le\; 2 \,.
\label{def_xc} 
\end{equation}
In this paper, let us assume that $n \in [0,2]$. The phase-transition line 
\eqref{def_xc} separates a diluted phase from a dense phase. 
The diluted phase $x < x_c(n)$ is, roughly speaking, a disordered phase with 
exponential decay of correlations.  
If $x\ge x_c(n)$, the model should be critical, with algebraic decay of 
correlations. In this case, the model is governed in the thermodynamic limit
by some CFT. Actually, there are two distinct critical regimes: $x=x_c(n)$ 
and $x > x_c(n)$, each of them with a distinct scaling limit. 
Figure~\ref{fig:1} shows the phase diagram of the O$(n)$ loop model in the
$(n,1/x)$ plane. The two branches $x_\pm$ \eqref{def_xplusminus} are
shown, as well as the dilute and dense phases for $n\le 2$. The phase
diagram for $n>2$ is beyond the scope of this paper.  

Although the partition functions of the O$(n)$ loop model \eqref{def_Z_loop} 
and the Potts model \eqref{def_Z_Potts}/\eqref{def_Z_FK} do not have a 
direct transformation, both can be represented in terms of a CG 
\cite{Nienhuis_87}. 
The RG analysis of the O$(n)$ loop model reveals that the RG trajectories
correspond to constant values of $n$, and that, for each $n\in [0,2]$, 
there are two types of fixed 
points (see Fig.~\ref{fig:1}) parametrized by \eqref{def_n_vs_g}.    
The $x_{+}=x_c$ branch corresponds to $1 \le g \le 2$, and 
represents unstable fixed points, indicating a second-order phase transition 
from the dilute phase to the dense phase (or the tricritical point of the 
Potts model). The $x_{-}$ branch (which is the analytic continuation of the
former branch) corresponds to $0 < g < 1$, and 
represents stable fixed points, reflecting the universality of the whole 
critical dense phase (or the ordinary critical point of the Potts model). 
We can then unify formulas \eqref{def_Q_vs_g}/\eqref{def_n_vs_g} as
\begin{equation}
  \sqrt{Q} \;=\; n \;=\; - 2\, \cos (\pi\, g)\,, \qquad g \;\in\; 
\left(0,2\right]\,. 
\label{def_nQ_vs_g}   
\end{equation} 
The physics of each of the points on $x_c$ is given by a CFT of central 
charge \cite{Cardy_87} 
\begin{equation}
c(g) \;=\; 1 - \frac{6\, (1-g)^2}{g}\,, \qquad g \;\in\; [1,2]\,. 
\label{def_c}
\end{equation}  
This equation can be analytically continued to cover also the critical Potts
model when $g\in (0,1]$. Note that the conformal charge of the dense 
phase at $n=\sqrt{Q}=\sqrt{2}$ or $g=3/4$ (the critical Ising model) is the
same as the conformal charge at $x_c(1)$ or $g=4/3$ (the one-state tricritical
Potts model).  

It is worth noticing that the critical value $x_c(0)=(2+\sqrt{2})^{-1/2}$ 
has been rigorously obtained in Ref.~\cite{Duminil-Copin_12}. 
On the other hand, at the point $(n,x) = (2,1/\sqrt{2})$, both branches 
$x_\pm$ merge. This point corresponds to the critical 4-state Potts model. 

As seen in the introduction, the leading thermal eigenvalue 
$y_{t1}^\text{loop}$ \eqref{def_y_t1_On} is not the leading thermal 
eigenvalue of the Potts model $y_{t1}$ \eqref{def_y_t1}, but the 
second leading eigenvalue $y_{t2}$ \eqref{def_y_t2}. 
As remarked in \cite{Deng_07}, the exponents $y_{t1}$ and $y_{h1}$ are 
\emph{absent} in spin and loop observables in the O$(n)$ model; but they can be
seen in observables associated to other objects, like faces.  
The O$(n)$ magnetic exponent \eqref{def_y_h1_On} also appears in connection to
an observable related to the worm MC algorithm \cite{Liu_11}. 

Finally, at $x=\infty$ or $1/x = 0$, there is another phase in the O$(n)$
loop model dubbed the fully packed phase. This phase is also critical, but 
its universality class is distinct from that of the dense phase 
\cite{Bloete_94,Batchelor_94,Kondev_96}. The physical properties of this
phase are beyond the scope of the present article.  

%
%
\section{\texorpdfstring{The correction-to-scaling exponent $\bm{\Omega}$}%
                        {The correction-to-scaling exponent Omega}}
\label{sec:Omega}

In this section, we will derive the conjecture \eqref{conj_Omega} for the 
correction-to-scaling exponent $\Omega$ [cf. Eq.~\eqref{def_ns_c}],
thereby extending Ziff's result for percolation $\Omega = 72/91$
\cite{Ziff_11}. As in Ziff's paper, we start from a formula due to
Saleur and Bauer \cite{Saleur_89} and rederived by
Cardy \cite{Cardy_06} that gives the partition function of the O$(n)$ model
on an annulus.  

Let us consider the annulus $0 \le x < \ell$ and $0 < y < L$ with periodic 
boundary conditions along the $x$ axis (i.e., we identify $x=0$ and $x=\ell$),
and free boundary conditions along the $y$ axis. 
In this section $\ell$ will denote the largest linear size of the annulus; 
we hope there is no confusion with the same letter denoting loop 
configurations in \eqref{def_Z_loop}/\eqref{def_Z_loop_Bis}.
We introduce the modulus
\begin{equation}
\widetilde{q} \;=\; e^{-2\pi L/\ell} \,.
\end{equation}
If we conformally map this region to an annulus of inner radius $R_1$
and outer radius $R$, then $\widetilde{q} = R_1/R$. 

The partition function of the O$(n)$ model on this annulus is given by 
\cite[formula on top of page 11]{Cardy_06} 
\begin{equation}
Z(\chi') \;=\; \frac{Z_\circ}{\sin\chi'} 
\sum\limits_{m\in\mathbb{Z}} a_m(\chi') \,
\widetilde{q}^{\,\frac{(\chi'+2\pi m)^2 - \chi^2}{2\pi^2 g}}
\label{def_Z_Cardy}
\end{equation}
where
\begin{subeqnarray}
Z_\circ &=& \sqrt{\frac{2}{g}}\, \widetilde{q}^{\,-\frac{c}{12}}\, 
\prod_{r=1}^\infty (1 - \widetilde{q}^{\,2r})^{-1} \,,
\slabel{def_Zo_Cardy} \\[2mm]
a_m(\chi') &=& \sin\left( \frac{\chi' + 2\pi m}{g}\right) 
\slabel{def_am_Cardy}   
\label{def_Zo_am_Cardy}
\end{subeqnarray}
Formulas \eqref{def_Z_Cardy}/\eqref{def_Zo_am_Cardy} correspond to a O$(n)$ 
model with CG coupling $g$, conformal charge $c=c(g)$ given by \eqref{def_c}
and parameter $\chi$. This parameter is related to $g$ by 
the standard formula
\begin{equation}
\chi \;=\; \pi (1-g) \,, 
\label{def_chi}
\end{equation}
and to the loop weight by $n=\sqrt{Q} = 2\, \cos \chi$. Again $g\in [1,2]$ 
corresponds to the tricritical $Q$-state Potts model, and $g\in(0,1]$
to the critical $Q$-state Potts model. 
The other parameter $\chi'$ corresponds to the weight given to 
noncontractible loops $n' = 2\, \cos\chi'$. Note that 
$Z_\circ$ \eqref{def_Zo_Cardy} is independent of $\chi'$.  

We now extract the terms independent of $m$ in the sum of 
Eq.~\eqref{def_Z_Cardy}:
\begin{equation}
Z(\chi') \;=\; \frac{Z_\circ}{\sin\chi'} \, 
\widetilde{q}^{\,\frac{\chi'\mbox{}^2-\chi^2}{2\pi^2 g} } \,  
\sum\limits_{m\in\mathbb{Z}} a_m(\chi') \,
\widetilde{q}^{\,\frac{2(\pi m^2 + \chi' m)}{\pi g}} \,.
\label{def_Z_Cardy_2}
\end{equation}

If we set $\cos\chi' = 0$ (i.e., $\chi'=\pi/2$), we suppress all contributions
with a nonzero number of noncontractible loops. This happens if
and only if there is a cluster connecting both boundaries at $y=0$ and
$y=L$. Therefore, the leading term of the crossing probability 
$\Pi(\widetilde{q})$ that there is a cluster connecting the boundaries is 
given in the limit $\widetilde{q}\to 0$ or $R_1/R\to 0$ by 
\begin{eqnarray}
\Pi(\widetilde{q}) \;=\;  \frac{Z(\pi/2)}{Z(\chi)} 
         &\sim& \widetilde{q}^{\, 1/(8g) - (1-g)^2/(2g)} \nonumber \\[2mm]
&=& 
               \left( \frac{R_1}{R} \right)^{1 - g/2 -3/(8g)} \,. 
\label{def_cross_P}
\end{eqnarray} 

The contributions from the sum in \eqref{def_Z_Cardy_2} when $\chi'=\pi/2$ is
given by
\begin{multline}
\sum\limits_{m\in\mathbb{Z}} a_m(\pi/2) \,
\widetilde{q}^{\, \frac{2(\pi m^2 + \pi m/2)}{\pi g}} \;=\; 
a_0 + a_{-1}\,  \widetilde{q}^{\, 1/g} \\[2mm]
+ a_1\, \widetilde{q}^{\, 3/g} +  \cdots 
\label{series1}
\end{multline} 
The contribution of the sum when $\chi'=\chi$ [cf. \eqref{def_chi}] is
\begin{multline}
\sum\limits_{m\in\mathbb{Z}} a_m(\chi) \,
\widetilde{q}^{\, \frac{2(\pi m^2 + \chi m)}{\pi g}} \;=\; 
a_0 + a_{-1}\,  \widetilde{q}^{\, 2} \\[2mm]
+ a_1\, \widetilde{q}^{\,4/g -2} +  \cdots 
\end{multline} 
As $g \in (0,2]$, both series generically have a nonzero leading term, 
and the most relevant correction term is $\widetilde{q}^{\, 1/g}$ from 
\eqref{series1}.

Therefore, the leading term of the crossing probability \eqref{def_cross_P}
is given by
\begin{eqnarray}
\Pi\left( \frac{R}{R_1} \right) &\sim& 
\left( \frac{R}{R_1} \right)^{ g/2 + 3/(8g) - 1} \big[ 1 + 
                     A \, \widetilde{q}^{\, 1/g} 
+ \cdots \big] \nonumber \quad\quad \\[2mm]
 &=& 
\left( \frac{R}{R_1} \right)^{d_f -2} \, \left[ 1 + A
       \left( \frac{R}{R_1} \right)^{-1/g} + \cdots \right] 
\label{def_cross_P_1} 
\end{eqnarray} 
where $d_f$ is the fractal dimension \eqref{def_df}.

Now we argue in the same way as in Ref.~\cite{Ziff_11}. The probability 
$\Pi(R/R_1)$ can be related to the probability at criticality 
$P_{\geq s}$ that an occupied vertex is connected to a cluster of size 
greater than or equal to $s$. This latter probability is given by
\begin{equation}
P_{\geq s} \;\sim\; s^{2-\tau} \, \big(1 + A\, s^{-\Omega} + \cdots \big)\,,
\label{def_Peqs} 
\end{equation} 
where $\tau$ is the Fisher exponent. 
At the critical temperature, the clusters are fractal objects, so their size 
$s$ and their radius $R$ are related as
\begin{equation}
 s \;\sim\; \left( \frac{R}{\epsilon} \right)^{d_f}\,,
\label{def_s_vs_s}
\end{equation} 
where $\epsilon$ is some length of the order of the lattice spacing,
and $d_f$ is the fractal dimension \eqref{def_df}. If we
assume that $\epsilon = R_1$, we have that the probability of having a cluster
of radius equal to or greater than $R$ is given by 
\begin{equation}
P_{\geq R} \;\sim\; R^{(2-\tau)\, d_f} \, \big(1 + A'\, R^{-\Omega \, d_f}
+ \cdots \big)\,.
\label{def_s_vs_R} 
\end{equation} 

Indeed, if we introduce the hyperscaling relation for $\tau$ \eqref{def_tau},
we have that  
\begin{equation}
(2-\tau) \, d_f \;=\; \left( 1 - \frac{2}{d_f} \right) \, d_f \;=\; 
d_f-2 \,, 
\end{equation} 
which is the leading exponent in \eqref{def_cross_P_1}. Therefore, the first
correction-to-scaling exponent should satisfy
\begin{equation}
\frac{1}{g} \;=\; \Omega \, d_f \,,
\end{equation}
which is the conjecture \eqref{conj_Omega}.

Note that the difference between the two most relevant magnetic eigenvalues
[cf.~Eq.~\eqref{def_y_h1_and_h2}] is precisely the leading 
correction-to-scaling term in \eqref{def_cross_P_1} 
\begin{equation}
y_{h2} - y_{h1} \;=\; -\frac{1}{g} \,.
\label{def_diff_y12}
\end{equation}

In other words, the fractal structure of the critical FK clusters should 
be characterized by both the leading and subleading magnetic exponents
\eqref{def_y_h1_and_h2}. If $R_s$ is the radius of a cluster of size $s$, 
one expects 
\begin{equation}
s \;\sim\; R_s^{y_{h1}}\,\left(1+b\, R_s^{y_{h2}-y_{h1}} + \cdots\right)\,.
\label{def_s_full}
\end{equation}

%
%
\section{Monte Carlo algorithm} \label{sec:MC}  

In this section, we discuss in more detail the MC algorithm used in 
the simulations of the O$(n)$ model on the hexagonal lattice performed in
this work to check conjecture \eqref{conj_Omega}. 

There are several MC algorithms that can be used to efficiently simulate
the O$(n)$ loop model on the hexagonal lattice: the worm algorithm 
\cite{Liu_11}, and extensions of the cluster Chayes--Machta 
\cite{Chayes_98} algorithm; see e.g., \cite{Deng_07,Fang_22}. Let us assume 
in this section that the graph $G$ is planar; e.g., a finite subset of the 
hexagonal lattice with free boundary conditions. We also consider integer 
values of $Q=n^2$.  

In the actual MC simulations, we employ the efficient cluster 
algorithm developed in Refs.~\cite{Deng_06,Fang_22}, which is based on 
the so-called induced-subgraph picture and was originally introduced by 
Chayes and Machta \cite{Chayes_98}.  

Basically, we start from the partition function of the model 
\eqref{def_Z_loop} and split the weight $n$ as follows:
\begin{equation}
n \;=\; n_\alpha + n_\beta \;=\; 1 + n_\beta \,,
\end{equation}
where $\alpha$ (resp.\ $\beta$) stands for the ``active'' (resp.\ ``inactive'')
color in the Chayes--Machta language. 
Now, given a loop configuration of the hexagonal lattice, we assign, 
independently for each loop, the active color $\alpha$ with probability $1/n$,
or the inactive color $\beta$ with probability $1-1/n$. So all vertices in
a given loop have the same color. In addition, we assign the active color
$\alpha$ to all vertices that do not belong to any loop. This is consistent
with the fact that these vertices have an implicit weight equal to one in
\eqref{def_Z_loop}. We have obtained an induced O$(1)$ (i.e., Ising)
model on the active vertices. This induced model can be simulated by any 
valid MC algorithm. 

It is important to realize that the loops in an O$(n)$ configuration on
the hexagonal lattice can be regarded as domain boundaries of an Ising model
of the dual triangular lattice; i.e., they represent the borders between
spin-up and spin-down regions. This correspondence allows us to map the
loop model on the hexagonal lattice $G$ to a generalized Ising model on the 
dual triangular lattice $G^*$. Note that this statement is true due to 
the planarity of $G$. Due to the spin-flip symmetry in the Ising model, there
is actually a one-to-two correspondence between loop configurations and 
dual Ising spin configurations. If $K^*$ is the nearest-neighbor coupling
for the dual Ising model~\footnote{%
  Please note that the Ising model is \emph{not} written as the 2-state Potts
  model \eqref{def_H_Potts}/\eqref{def_Z_Potts}. We use instead the standard 
  Ising model notation given by Eq.~\eqref{def_Z_GIsing}. 
}, the bond weight $x$ is given by  
\begin{equation}
2\, K^* \;=\; - \log(x) \,.
\end{equation}
Indeed, when $0<x<1$, the Ising model is ferromagnetic (for $x>1$, the system 
is antiferromagnetic). The partition function of this generalized Ising model 
on the dual triangular lattice $G^* = (V^*,E^*)$ is given by
\begin{equation}
Z_\text{GIsing}(G^*;K^*,n) \;=\; \sum\limits_{\{s\}} n^{\mathcal{N}_d} \, 
                      \prod_{\{i,j\}\in E^*} e^{K^*\, s_i s_j} \,,
\label{def_Z_GIsing}
\end{equation}
where $\mathcal{N}_d = \mathcal{N}_\ell + 1$ is the number of spin 
domains (within each spin domain, spins are identical), and the summation 
is over all Ising configurations. The relation between the O$(n)$ model
on the hexagonal lattice and the generalized Ising model \eqref{def_Z_GIsing}
on the dual triangular lattice is shown in Fig.~\ref{fig:2}. The O$(n)$ loops
live on the edges of the original hexagonal lattice, while the Ising spins 
live on the vertices of the dual triangular lattice. The value of these Ising
spins is given by the color of the corresponding face. 

We can now apply the induced subgraph method to the generalized Ising model 
\eqref{def_Z_GIsing} on $G^*$. From any given spin configuration, we compute 
the corresponding spin domains. Then, independently for each spin domain, 
we assign the active color $\alpha$ (resp.\ inactive color $\beta$) 
with probability $1/n$ (resp.\ $1-1/n$). 
In this way, we obtain a random partition of the lattice $G^*=(V^*,E^*)$. 
The vertex set is naturally split as $V^* = V^*_\alpha \cup V^*_\beta$. 
The edge set is also partitioned into 
$E^* = E^*_\alpha \cup E^*_\beta \cup E^*_{\alpha\beta}$. Here, $E^*_a$ (with
$a=\alpha,\beta$) is the subset of edges whose endpoints both belong to 
$V^*_a$. The set $E^*_{\alpha,\beta}$ contains all edges such that one vertex
belongs to $V^*_\alpha$, and the other to $V^*_\beta$. 

Note that we have an induced Ising model on the active subgraph $G^*_\alpha$; 
this model can be simulated using the SW algorithm \cite{Swendsen_87}. The 
probability $p$ that a bond is occupied is chosen, when the system is 
ferromagnetic ($0<x<1$), as $p=1-e^{-2K^*}=1-x$ for each pair of 
parallel neighboring spins. If the system
is antiferromagnetic ($x>1$), the bond occupation probability is 
$p=1-e^{2K^*}=1-1/x$ for each pair of anti-parallel neighboring spins. 
This step needs some extra care: all edges not
belonging to $E^*_\alpha$ must be occupied to ensure that the domain topology 
of the inactive subgraph $G^*\setminus G^*_\alpha$ remains unchanged.  

In practice, we are going to simulate the generalized Ising model 
\eqref{def_Z_GIsing} on finite triangular lattices of size $L\times L$ with 
periodic boundary conditions. This model is not, strictly speaking, identical
to the O$(n)$ loop model on a finite hexagonal lattice with periodic boundary 
conditions. Nevertheless, such a boundary difference is expected to play a 
negligible role in the bulk critical phenomena, and indeed, no practical effect 
has been observed in previous studies \cite{Fang_22}.   

The detailed procedure is as follows:

\begin{enumerate}

\item[(1)] Independently for each spin domain, choose the active color 
$\alpha$ with probability $1/n$, or the inactive color $\beta$ with 
probability $1-1/n$. 

\item[(2)] Independently for each edge $e \in E^*$, place an occupied bond
with probability $p$. For the ferromagnetic case $x\in (0,1)$, this probability 
is given by  
\begin{equation}
p\;=\; \begin{cases}
  1-x  &   \text{if $s_i \;=\; s_j$, and $e\in E^*_{\alpha}$,} \\
    0  &   \text{if $s_i \;\neq\; s_j$, and $e\in E^*_{\alpha}$,} \\
    1  &   \text{otherwise.}
\end{cases}
\label{def_p_FM}
\end{equation}
For the antiferromagnetic case $x>1$, we have
\begin{equation}
p\;=\; \begin{cases}
  1-\frac{1}{x}  &   \text{if $s_i \;\neq\; s_j$, and $e\in E^*_{\alpha}$,} \\
    0  &   \text{if $s_i \;=\; s_j$, and $e\in E^*_{\alpha}$,} \\
    1  &   \text{otherwise.}
\end{cases}
\label{def_p_AF}
\end{equation}
  
\item[(3)] We now obtain a cluster configuration on $G^*_\alpha$. 
Note that, even for the ferromagnetic case, a cluster typically contains 
both $+1$ and $-1$ Ising spins. 
Independently, for each connected component, flip the Ising spins or 
keep the current Ising spins with probability $1/2$.
The new loop configuration is the Peierls contour for the new 
Ising configuration.
\end{enumerate} 

%
%
\begin{figure*}[tbh]
\centering
\begin{tabular}{cccc}
\multicolumn{4}{c}{\includegraphics[scale=0.95]{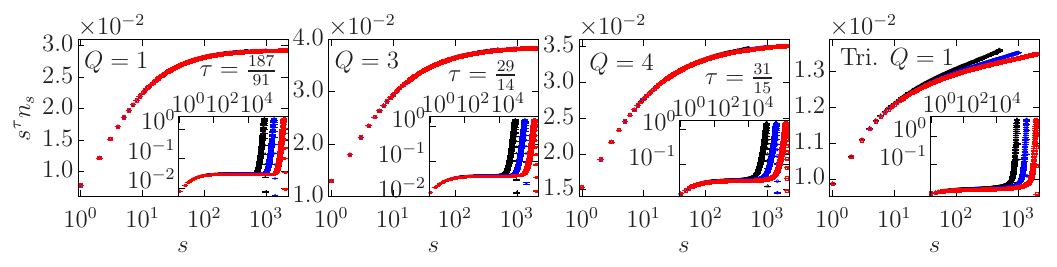}} \\
 \hspace{0.148\textwidth} (a) & 
 \hspace{0.185\textwidth} (b) & 
 \hspace{0.190\textwidth} (c) &  
 \hspace{0.093\textwidth} (d) \\
\end{tabular}
\caption{\label{fig:3}
Finite-cluster-size effects on $n_s$ for the critical and tricritical 
Potts models. 
We plot the modified size distribution $s^{\tau}\, n_s$ of FK clusters versus 
$s$ (with $\tau$ fixed at it exact value), in the critical $Q=1$ (a), 
$Q=3$ (b), and $Q=4$ (c) Potts models, as well as in the tricritical $Q=1$ (d) 
Potts model. The black, blue, and red data points correspond to systems of 
linear sizes $L=512, 1024$ and $2048$, respectively. 
In the main figure of each panel, we only display data with $s\leq L$, 
which has no wrapping effect.
The excellent data collapse for the various system sizes in panels (a)--(c)
shows that $n(s)$ on the $x_{-}$ branch has no noticeable finite-lattice-size 
corrections. 
On the other hand, the main figure in panel (d) shows sizable 
finite-lattice-size corrections. The Fisher exponent for the tricritical 
one-state Potts model (d) is $\tau=379/187$.
In each panel, the inset plots $s^\tau\, n_s$ for all values of $s$, 
demonstrating that the wrapping effect from periodic boundary 
conditions always exists.}
\end{figure*}

It is known \cite{Fang_22} that the above algorithm for the O$(n)$ loop model
in the dense phase has no CSD. Along the dilute critical branch, the CSD is 
also completely absent for $1 < n <2$;  for $n=1$, the algorithm just reduces 
to the standard SW cluster method, which is known to have some minor CSD
$z \approx 0.22$ \cite{Salas_00}. 
In addition, since in this project we are 
interested in the geometric structure of the Ising domains, simulating the
generalized Ising model \eqref{def_Z_GIsing} is also very convenient, as we
do not have to go back and forth between the hexagonal and the 
triangular lattices.

%
%
\section{Results}  \label{sec:res}
	
We have simulated the generalized Ising model \eqref{def_Z_GIsing} on 
triangular lattices of size $L\times L$ with $L=16,32,\ldots,2048$, and 
periodic boundary conditions.
We have selected the values $n = 1, \sqrt{2}, \sqrt{3}$ (i.e., $Q=1,2,3$)
along the $x_{-}$ branch (\ref{def_xplusminus}, with negative sign),  
and $n = 1, \sqrt{2}$ along the $x_{+}$ branch (\ref{def_xplusminus}, with
positive sign),  
as well as $n = 2$ (i.e., $Q=4$) where $x_\pm$ meet. 
For each system, we have generated over $10^6$ statistically 
independent samples.	

Studies on correction-to-scaling exponents in the percolation model, both 
numerical and theoretical, are typically conducted in infinite systems. 
In finite systems of linear size $L$, the critical divergent correlation 
length $\xi$ is smeared out to be of order $O(L)$, and the size of the 
largest cluster scales as $\sim L^{d_f}$. As a result, the asymptotic 
cluster-size distribution $n_s \sim s^{-\tau}$ is modified to have a 
finite-lattice-size scaling form as 
\begin{equation}
n_s \;\sim\; s^{-\tau}\, \widetilde{n}\left(s\, L^{-d_f}\right)\,,
\end{equation} 
where $\widetilde{n}(x)$ is a universal function and, typically, decays 
exponentially for $x \gg 1$. To study the cluster-size distribution and 
particularly the finite-cluster-size corrections, one should restrict to 
the region $1 \ll s \ll L^{d_f}$. In particular, to avoid any wrapping 
effects from the periodic boundary conditions, it would be even better to 
explore in the more restricted region $ 1 \ll s \leq L$.

Figure~\ref{fig:3} illustrates the impact of finite-lattice-size effects and
boundary conditions on the observable 
$n_s$ by plotting $s^{\tau}\, n_s$ versus $s$ for points on the two branches
$x_\pm$ \eqref{def_xplusminus}. 
Figures \ref{fig:3}(a) and \ref{fig:3}(b) 
represent data on the $x_-$ branch: (a) $n=1$, and 
(b) $n=\sqrt{3}$, respectively. The main figure in these panels shows MC data 
for $L=512$, 1024, and 2048 up to $s=L$; i.e., the linear size $L$ is used as 
an upper cutoff. Moreover, the corresponding insets show the full data set.  
It can be observed in the insets that 
data points for $s\gtrsim L$ are clearly affected by finite-lattice-size 
effects. That is why we have used the system
linear size $L$ as an upper cutoff in our analysis.  
As discussed in the Introduction, RG studies indicate that for the
O$(n)$ loop model on the $x_-$ branch with $1\le n \le 2$, the subleading 
eigenvalue $y_{t2}$ becomes irrelevant. Therefore, its contribution to 
finite-lattice-size effects is negligible for all practical purposes. 
In panels (a) and (b), it is clear that 
the data for different system sizes fall on the same curve when $L$ is used 
as a cutoff, indicating no system-size dependence. We can safely assume that
we are working in the thermodynamic limit.  
Similar observations apply to panel (c), which corresponds to $n=2$. At this
point both branches $x_\pm$ meet. However, we can see small finite-lattice-size 
effects for the $L=512$ data. Finally, the rightmost panel (d) displays 
the data for $n=1$ on the branch $x_+$. In the main figure, we can observe 
a noticeable separation of the data points for $s\gtrsim 10^2$. This is a 
sign of significant finite-lattice-size effects, which make harder the data 
analysis for this branch.  

%
%
\subsection{\texorpdfstring{Critical Potts model with $\bm{Q=1,2,3}$}%
                           {Critical Potts model with Q=1,2,3}}
\label{sec:res_x_minus}

In this section we will consider three points on the branch $x_-$, namely
$Q=1,2,3$ (or, equivalently, $n=1,\sqrt{2},\sqrt{3}$). These points 
belong to the universality class of the critical $Q$-state Potts model. 

The size distribution of FK clusters at the critical point follows the 
behavior $n_s = s^{-\tau}\, (A + B\, s^{-\Omega} + \cdots)$ 
for large enough values of $s$ [cf.~\eqref{def_ns_c}]. 
The parameter $\tau$ is the Fisher exponent \eqref{def_tau}. 
We have employed a least-squares fit to the data using the transformed variable 
\begin{equation}
s^{\tau} \, n_s \;=\;  a + b_1\, s^{-y_1}\,,
\label{def_Ansatz_ns}
\end{equation}
and performed a systematic analysis by applying different lower cutoffs
$s\ge s_\text{min}$. 
For fitting the data on the $x_{-}$ branch, we have only used data from the 
largest system $L = 2048$, as the data for the smaller system sizes 
coincide perfectly for $s\ge 128$. 
We have also attempted to include an additional correction term 
$b_2 s^{-y_2}$ (with $y_2 > y_1 >0$) in the ansatz \eqref{def_Ansatz_ns}, 
but this did not result in smaller error bars for the final estimates. 
In Table~\ref{tab:fit1}, we 
summarized the results of the fits. For each value of $Q=1,2,3$, we 
show several fits corresponding to distinct choices of the lower cutoff 
$s_\text{min}$. For each fit, we display the parameters $a, b_1$, and $y_1$
[cf.~\eqref{def_Ansatz_ns}], as well as the $\chi^2$ and the 
number of degrees of freedom (DF) of the corresponding fit. 
When the cutoff $s_\text{min}$ is too small, finite-size effects at small 
$s$ cause a significant increase in $\chi^2/\text{DF}$. 

%
%
\begin{table}[tbh]
\caption{\label{tab:fit1}
Power-law fits \eqref{def_Ansatz_ns} to the Monte Carlo data for the
critical Potts model with $Q=1, 2, 3$ states. For each value of $Q$, we
provide two distinct fits, each of them corresponding to a different
lower cutoff $s\ge s_\text{min}$ with $s_\text{min}=64, 96, 128$. 
For each single fit, we provide the estimates of the free parameters 
$a, b_1, y_1$, as well as the values of the $\chi^2$ and the number of 
degrees of freedom (DF) of the fit. The entries with an exact number for
$y_1$ correspond to the two-parameter fits to the ansatz 
\eqref{def_Ansatz_ns} where $y_1$ is fixed to the conjectured value for
$\Omega$ \eqref{conj_Omega}.}
\begin{ruledtabular}
\begin{tabular}{crllll}
Q   & $s_{\rm min}$ & \multicolumn{1}{c}{$y_1$} 
                    & \multicolumn{1}{c}{$b_1$} 
                    & \multicolumn{1}{c}{$a$} 
                    & \multicolumn{1}{c}{$\chi^2/{\rm DF}$} \\[1mm]
\hline \\[-3mm] 
1   & 64  &0.790(2) &-0.0461(3)    &0.029351(2)    &328.6/317\\
    & 96  &0.796(3) &-0.0472(6)    &0.029347(2)    &281.4/285\\
    & 128 &0.795(5) &-0.047(1)     &0.029347(3)    &251.7/253\\
    & 96  & 72/91   &-0.04639(4)   &0.0293501(8)   &283.4/286\\
    & 128 & 72/91   &-0.04633(6)   &0.0293495(9)   &252.3/254\\
    & 160 & 72/91   &-0.04621(9)   &0.029348(1)    &237.7/238\\[1mm]
\hline \\[-3mm]
2   & 64  &0.716(2) &-0.0480(3)     &0.036014(3)    &336.9/317\\
    & 96  &0.723(4) &-0.0494(7)     &0.036006(5)    &307.1/285\\
    & 128 &0.723(6) &-0.049(1)      &0.036006(6)    &267.3/253\\
    & 96  &32/45    &-0.04714(5)    &0.036021(1)    &318.2/286\\
    & 128 &32/45    &-0.04699(8)    &0.036018(2)    &272.1/254\\
    & 160 &32/45    &-0.0469(1)     &0.036016(2)    &246.2/238\\[1mm]
\hline \\[-3mm]
3   & 64  &0.647(2) &-0.0430(3)     &0.038649(5)    &324.7/317\\
    & 96  & 0.647(4)&-0.0430(6)     &0.038649(7)    &303.4/285\\
    & 128 & 0.639(6)&-0.0415(10)    &0.038660(9)    &267.9/253\\
    & 96  & 9/14    &-0.04228(5)    &0.038657(2)    &304.8/286\\
    & 128 & 9/14    &-0.04215(7)    &0.038654(2)    &268.4/254\\
    & 160 & 9/14    &-0.0421(1)     &0.038654(3)    &257.7/238\\[1mm]
\end{tabular}
\end{ruledtabular}
\end{table}

%
%
\begin{figure}[tbh]
\centering
\includegraphics[scale=0.9]{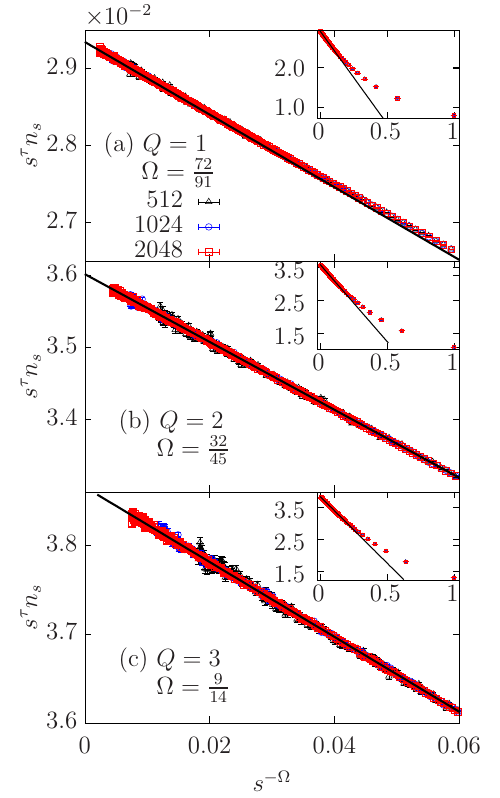}
\caption{\label{fig:4}
Estimates of the correction-to-scaling exponent $\Omega$ for the critical 
Potts model with $Q=1$ (a), $Q=2$ (b), and $Q=3$ (c). Each main 
figure depicts $s^{\tau}\, n(s)$ versus $s^{-\Omega}$ for various system sizes
$L=512$ (black), $L=1024$ (blue), and $L=2048$ (red). In these plots, 
$\Omega$ is fixed to its conjectured value \eqref{conj_Omega}, which is 
depicted in each panel. 
In each main plot we show the data with $s \gtrsim 35$.  
The black lines correspond to the two-parameter fit to the ansatz
\eqref{def_Ansatz_ns} with $\Omega$ fixed to its conjectured value. 
In the insets, we show the full data set to highlight that, at small values
of $s$, the behavior is not linear.
}
\end{figure}

The point $n=1$ on the $x_-$ branch corresponds to $g=2/3$ and 
$\tau = 187/91 \approx 2.054\,945$. This is the bond-percolation critical 
point,
for which the correction-to-scaling exponent has been precisely 
determined as $\Omega=72/91 \approx 0.791\,209$ \cite{Ziff_11}. 
This value is also predicted by conjecture \eqref{conj_Omega}. 
We see in Table~\ref{tab:fit1} that the results for $s_\text{min}=64, 128$ 
are compatible within errors, giving $\Omega = 0.795(5)$, which fully agrees
with the above value. 
In Fig.~\ref{fig:4}(a), we display the data for 
$s^{\tau} n_s$ vs $s^{-\Omega}$ with $\Omega=72/91$. The agreement is very
good for large enough values of $s$. In the inset, we show the full data set,
and we see that at small values of $s$, the behavior of 
$s^{\tau} n_s$ vs $s^{-\Omega}$ is no longer linear, as expected. In this
regime, additional correction-to-scaling terms should contribute, and the
simple ansatz \eqref{def_Ansatz_ns} is not valid any more.  

The point $n=\sqrt{2}$ on the $x_-$ branch corresponds to $g=3/4$ and 
$\tau = 31/15 \approx 2.066\,667$. This is the critical point of the  
Ising model. Conjecture \eqref{conj_Omega} predicts 
$\Omega = 32/45 \approx 0.711\,111$. 
We see in Table~\ref{tab:fit1} that the results for $s_\text{min}=64, 128$ 
are compatible within errors, giving $\Omega = 0.723(6)$, which agrees
within two standard deviations from the conjectured value. 
In Fig.~\ref{fig:4}(b), we display the data for 
$s^{\tau} n_s$ vs $s^{-\Omega}$ with $\Omega=32/45$. Again the agreement
for large $s$ is very good; while at small values of $s$, we need more
terms in the ansatz \eqref{def_Ansatz_ns}.  

The point $n=\sqrt{3}$ on the $x_-$ branch corresponds to $g=5/6$ and 
$\tau = 29/14 \approx 2.071\,429$. This is the critical point of the  
3-state Potts model. Conjecture \eqref{conj_Omega} predicts 
$\Omega = 9/14 \approx 0.642\,857$. 
We see in Table~\ref{tab:fit1} that the results for $s_\text{min}=64,96$ 
are compatible within errors, and they differ by two standard deviations from
the results with $s_\text{min}=128$. This differences can be explained by
systematic errors not taken into account in the ansatz \eqref{def_Ansatz_ns}. 
Our preferred estimate would be the one
for $s_\text{min}=128$ with some conservative error bars to take into
account the behavior of the estimates versus $s_\text{min}$: 
$\Omega = 0.638(9)$. The agreement with the conjectured value is very good. 
In Fig.~\ref{fig:4}(c), we display the data for 
$s^{\tau} n_s$ vs $s^{-\Omega}$ with $\Omega=9/14$. The same comments made
for the other two values of $Q$ apply in this case, too.  

%
%
\begin{figure}[bth]
\centering
\includegraphics[scale=0.9]{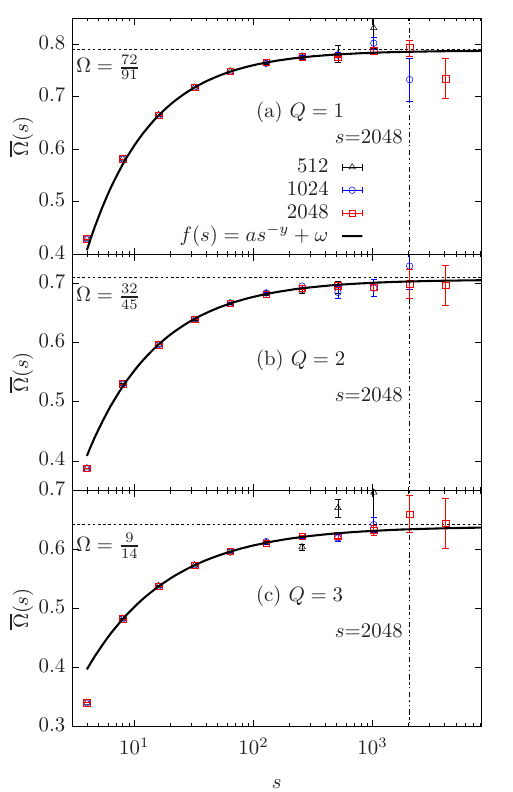}
\caption{\label{fig:5}
Quantity $\overline{\Omega}$ \eqref{def_bar_Omega} for the critical Potts 
model. We show the plots of $\overline{\Omega}(s)$ vs $s$ to show 
our conjectured correction-to-scaling exponent $\Omega$ for the critical 
Potts model with $Q = 1$ (a), $2$ (b), and $3$ (c). Each figure plots 
$\overline{\Omega}(s)$ for three various system sizes $L=512, 1024, 2048$ by 
choosing $s=4,8,\dots,2L$. The vertical dashed lines 
indicate $s=2048$, which 
is the linear system size for $L=2048$ data points. Note that for
$s > L$, the error bars for the data are significantly larger than  
those with $s\le L$. 
The horizontal dashed lines represent our conjectured correction-to-scaling 
exponent $\Omega$ \eqref{conj_Omega}. The black curves represent fits to the 
$L=2048$ data (red points) to the ansatz \eqref{def_Ansatz_bar_Omega}, 
excluding the $s \geq L$ data points.
}
\end{figure}   

Ziff proposed an intuitive method to observe the value of $\Omega$ in 
the percolation model \cite{Ziff_11}, which we have extended to the Potts 
model. The idea was to measure the probability $P_{\geq s}$ that an occupied 
site belongs to an FK cluster of size greater than or equal to $s$. Its 
behavior is similar to that in the percolation model
\begin{equation}
P_{\geq s} \;=\; s^{2-\tau}\, \Big( A + B \, s^{-\Omega}+\cdots\Big)\,.
\end{equation}
The value of $\Omega$ can be directly assessed through
\begin{equation}
\overline{\Omega}(s) \;=\; 
-\text{log}_{2} \left(\frac{C_{s}-C_{s/2}}{C_{s/2}-C_{s/4}}\right)
\label{def_bar_Omega}
\end{equation}
where 
\begin{equation}
C_s \;=\; s^{\tau-2}\, P_{\geq s}\,.
\end{equation}
The leading order of the linear combination $C_{s}-C_{s/2}$ equals 
$B (1-2^{\Omega})\, s^{-\Omega}$; this implies formula \eqref{def_bar_Omega}. 
Due to higher-order correction-to-scaling effects, $\overline{\Omega}$ 
deviates from $\Omega$ at small $s$, but gradually approaches to $\Omega$ 
as $s$ increases.

Figures~\ref{fig:5}(a)--\ref{fig:5}(c) 
show the plots $\overline{\Omega}$ vs $s$ for the critical 
Potts model with $Q=1$, $2$, and $3$, respectively. The vertical dashed line 
represents $s=2048$, corresponding to the upper cutoff to the data point
with $L=2048$ (red points). It can be observed that the error bars 
significantly increase when $s \geq L$, consistent with the previous 
truncation of $n_s$ data. We have fitted the data to a power-law ansatz
\begin{equation}
\overline{\Omega}(s) \;=\; \omega + a\, s^{-y}\,, \quad y\;>\; 0 \,.
\label{def_Ansatz_bar_Omega}
\end{equation}
The second tern in this ansatz takes into account higher-order corrections.
We have included in the fits the data for $\overline{\Omega}$ with  
$s < s_{\text{max}} = L$. The horizontal dashed lines in Fig.~\ref{fig:5}
indicate the conjectured values for $\Omega$, and the gradual approach of 
$\overline{\Omega}$ to the conjectured value of $\Omega$ as $s$ increases 
further supports our ansatz \eqref{def_Ansatz_bar_Omega}. 

The critical hexagonal-lattice O$(1)$ loop model in the dense phase 
is equivalent to site percolation on the dual triangular lattice. 
Ziff has already studied the properties of this latter system in the 
thermodynamic limit \cite{Ziff_11}. 
We plotted $\overline{\Omega}$ versus $s$ for various system sizes in
Fig.~\ref{fig:5}(a). The behavior is consistent with Ziff's results. 
We then fitted the $\overline{\Omega}$ data to the ansatz
\eqref{def_Ansatz_bar_Omega}, excluding the points with $s \geq L$ to eliminate 
the effects of wrapping clusters. The estimates from the fit yielded 
$a=-1.17(4)$, $y=0.81(2)$, and $\omega=0.788(2)$. This result differ by $1.6$
standard deviations from the exact result $\Omega=72/91$. 

To deal with the critical Potts model with $Q=2$ and $Q=3$, we 
have used a similar method to fit the data for $s \le L$. 
The results for $Q=2$ of the fits to the ansatz 
\eqref{def_Ansatz_bar_Omega} are $a=-0.81(2)$, $y=0.72(1)$, and 
$\omega=0.707(3)$. This estimate differs from the exact value $\Omega=32/45$
in $1.4$ standard deviations [see Fig.~\ref{fig:5}(b)]. 
For $Q=3$, we have $a=-0.57(2)$, $y=0.62(2)$, and $\omega=0.639(3)$. In this
case, the latter estimate is $1.3$ standard deviations away from the exact 
value $\Omega=9/14$ [see Fig.~\ref{fig:5}(c)]. 
It can be seen in Figs.~\ref{fig:5}(b)--(c) that the data points approach the
expected value in a similar way as in Fig.~\ref{fig:5}(a).

\subsection{\texorpdfstring{Tricritical Potts model with
    $\bm{Q=1,2,3}$ and $\bm{Q=4}$}%
   {Tricritical Potts model with Q=1,2 and Q=4}} \label{sec:res_x_plus} 
	
In this section we consider two points on the branch $x_+$, namely
$Q=1,2$ (or, equivalently, $n=1,\sqrt{2}$). These points 
belong to the universality class of the tricritical $Q$-state Potts model. 
In addition, we consider the critical Potts with
$Q=4$ states (or, equivalently, $n=2$). At this point, both branches 
$x_\pm$ merge, and sophisticated finite-lattice-size corrections
might occur. Note that the case $Q=3$ has not been studied, as the results 
should behave in a similar way as those for $Q=2,4$. 

%
%
\begin{figure}[tbh]
\centering
\includegraphics[scale=0.9]{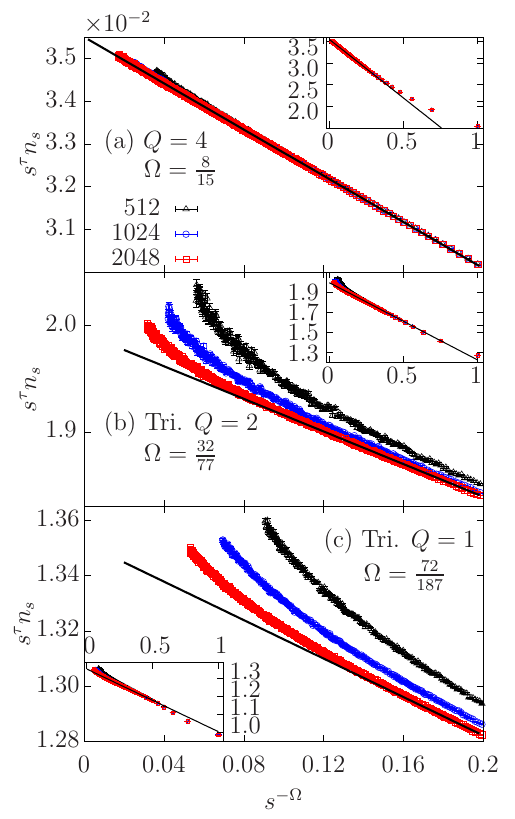}
\caption{\label{fig:6}
Estimates of the correction-to-scaling exponent $\Omega$ for the critical 
Potts model with $Q=4$ (a) and for the tricritical Potts model with 
$Q=2$ (b), and $Q=1$ (c). Each main 
figure depicts $s^{\tau}\, n(s)$ versus $s^{-\Omega}$ for various system sizes
$L=512$ (black), $L=1024$ (blue), and $L=2048$ (red). In these plots, 
$\Omega$ is fixed to its conjectured value \eqref{conj_Omega}, which is 
depicted in each panel.  
The black lines correspond to the two-parameter fit to the ansatz
\eqref{def_Ansatz_ns} with $\Omega$ fixed to its conjectured value. 
In each main plot we show the data with $s \gtrsim 21$.  
In the insets, we show the full data set to highlight the effects at large and
small values of $s$.
}
\end{figure}	

%
%
\begin{table}[tbh]
\caption{\label{tab:fit2}
Power-law fits \eqref{def_Ansatz_ns} to the Monte Carlo data for the
critical Potts model with $Q=4$ states and the tricritical Potts model
with $Q=1,2$ states. For each value of $Q$, we
provide several distinct fits, each corresponding to a different
lower and upper cutoffs $s_\text{max} \ge s\ge s_\text{min}$. 
For each single fit, we provide the estimates of the free parameters 
$a, b_1, y_1$, as well as the values of the residual $\chi^2$ and the number of 
degrees of freedom (DF) of the fit. The entries with an exact number for
$y_1$ correspond to the two-parameter fits to the ansatz
\eqref{def_Ansatz_ns} where $y_1$ is fixed to the conjectured value for
$\Omega$ \eqref{conj_Omega}. 
}
\begin{ruledtabular}
\begin{tabular}{cllllll}
Q   & $s_{\rm min}$ & $s_{\rm max}$ 
                    & \multicolumn{1}{c}{$y_1$}
                    & \multicolumn{1}{c}{$b_1 \times 10^2$}
                    & \multicolumn{1}{c}{$a \times 10^2$}
                    & \multicolumn{1}{c}{$\chi^2/{\rm DF}$} \\[1mm]
\hline \\[-3mm]
4 & 32 & 512 &0.5319(7) &-2.702(5)    &3.5511(3)    &259.7/221\\
  & 64 & 512 &0.536(2)  &-2.73(2)     &3.5496(8)    &212.8/189\\
  & 32 & 640 &0.5315(7) &-2.700(5)    &3.5513(3)    &275.0/237\\
  & 64 & 640 &0.534(2)  &-2.72(2)     &3.5505(6)    &231.5/205\\
  & 32 & 768 &0.5313(6) &-2.698(4)    &3.5514(3)    &290.0/253\\
  & 64 & 768 &0.532(2)  &-2.71(1)     &3.5510(6)    &248.3/221\\
  & 32 & 640 &8/15      &-2.7118(6)   &3.55045(7)   &283.5/238\\
  & 64 & 640 &8/15      &-2.714(1)    &3.5506(1)    &231.6/206\\
  & 96 & 640 &8/15      &-2.716(2)    &3.5507(2)    &191.5/174\\ [1mm]
\hline \\[-3mm]
Tri.~2 
  & 16 & 96  &0.419(2)  &-0.767(2)    &1.9907(8)    &70.0/78\\
  & 32 & 96  &0.415(8)  &-0.764(9)    &1.992(3)     &48.2/62  \\
  & 16 & 112 &0.417(2)  &-0.766(1)    &1.9917(7)    &91.2/94 \\
  & 32 & 112 &0.409(6)  &-0.758(7)    &1.995(2)     &65.9/78  \\
  & 16 & 128 &0.416(1)  &-0.765(1)    &1.9921(6)    &106.4/109\\
  & 32 & 128 &0.407(5)  &-0.756(6)    &1.995(2)     &79.0/93  \\
  & 16 & 112 &32/77     &-0.7649(3)   &1.99210(8)   &91.6/95\\
  & 32 & 112 &32/77     &-0.7655(8)   &1.9922(2)    &66.9/79 \\
  & 48 & 112 &32/77     &-0.767(2)    &1.9924(3)    &53.3/63 \\[1mm]
\hline \\[-3mm]
Tri.~1 
  & 24 & 68  &0.372(4)  &-0.344(1)    &1.355(1)     &41.5/42\\
  & 32 & 68  &0.37(1)   &-0.345(4)    &1.355(2)     &31.4/34\\
  & 24 & 68  &72/187    &-0.3486(2)   &1.35174(6)   &50.8/43\\
  & 32 & 68  &72/187    &-0.3494(4)   &1.35193(8)   &32.8/35\\[1mm]
\end{tabular}
\end{ruledtabular}
\end{table}

The tricritical Potts model corresponds to the unstable fixed 
point of the O($n$) loop model. We have assumes that the size distribution 
$n_s$ of the FK clusters exhibits the same behavior as in the critical 
Potts model \eqref{def_ns_c}, and that our conjectured formula for the 
correction-to-scaling exponent is given by formula \eqref{conj_Omega}. 
In contrast to the dense branch $x_-$, the subleading thermal scaling field 
is relevant  ($y_{t2}>0$) along the dilute branch $x_+$. As a consequence, 
the thermal fluctuations become significant, contributing as a source of 
finite-lattice-size corrections, and becoming more and more severe as 
$n$ decreases (i,e., as we approach $n=Q=1$).

The above argument implies, in practice, that a direct fit of
the $n_s$ data to the ansatz \eqref{def_Ansatz_ns} does not yield stable
results. If we plot $s^{\tau}n_s$ vs $s^{-\Omega}$ using the conjectured
value for $\Omega$ \eqref{conj_Omega}, we obtain Fig.~\ref{fig:6}. 
In the insets of the panels, we observe that the behavior is not linear when
$s$ is small (as in Sec.~\ref{sec:res_x_minus}), but also for large enough
values of $s$, as it can be seen in the main panels of Fig.~\ref{fig:6}. 
In this figure, we have depicted data for three the values $L=512,1024$, 
and 2048.
This latter effect is due to strong finite-size-scaling corrections. 
Indeed, as $L$ grows, the behavior of the corresponding data becomes linear
in a wider interval of $s$.
In order to isolate the linear regime of $s^{\tau}n_s$ vs $s^{-\Omega}$,
we have consider, for each value of $L$, the data in the interval 
$[s_\text{min},s_\text{max}]$ for some values of $s_\text{min}<s_\text{max}$.
Then, by varying systematically both $s_{\rm max}$ and $s_{\rm min}$, we 
fitted the data to the ansatz \eqref{def_Ansatz_ns}, and in this 
case we obtained stable results for the three values of $L$ we 
considered in this analysis.
The solid lines in Fig.~\ref{fig:6} represent the best fit for the
largest-size data ($L=2048$).
In Table~\ref{tab:fit2}, we 
summarized the results of the fits. For each value of $Q=4,2,1$, we 
show several fits corresponding to distinct choices of the lower and upper
cutoffs $s_\text{min},s_\text{max}$. For each fit, we display the 
parameters $a, b_1$, and $y_1$
[cf.~\eqref{def_Ansatz_ns}], as well as the $\chi^2$ and the 
number of degrees of freedom (DF) of the corresponding fit. 

The four-state critical Potts model ($g=1$) has $\tau=31/15$ 
and we conjecture $\Omega=8/15 \approx 0.533\,333$. 
The data presented in Fig.~\ref{fig:6}(a) shows a good agreement with a 
straight line in the region where $s \gg 1$. We also observe slight 
deviations from linearity for $s \gtrsim L$. We performed a systematic 
fit of the data to the power-law ansatz \eqref{def_Ansatz_ns} by setting 
different upper and lower cutoffs $s_\text{min}$ and $s_\text{max}$. First, 
we selected the value of $s_{\rm max}\gtrsim L$ from Fig.~\ref{fig:6}(a)
where no significant deviation from linearity is observed. We then 
progressively chose different values of $s_{\rm min}$ in increasing order,
starting from $s_\text{min}=16$. If the variance of the
residuals decreases when we increase $s_{\rm min}$, and the estimates for  
the parameters $y_1$, $a$, and $b_1$ are stable within errors, then 
$s_{\rm max}$ is considered a reasonable choice. 
Subsequently, $s_{\rm max}$ is reduced, and the procedure is repeated to 
ensure the reliability of the fit. Some of the results are shown in 
Table~\ref{tab:fit2}, where the fit is relatively stable for $s \leq 768$, 
yielding $\Omega=0.534(4)$, which agrees very well with the conjecture.

The 2-state tricritical Potts model has $g=5/4$, $\tau=157/77$, and we
conjecture $\Omega=33/77 \approx 0.428\,571$. 
Following the same procedure as for $Q=4$, we first plotted 
$s^{\tau}n_s$ versus $s$ in Fig.~\ref{fig:6}(b) using the conjectured value 
for $\Omega$. As discussed earlier, if we are on the branch $x_+$ and
the parameter $n$ is decreased, the finite-size effects from $L$ 
become larger. 
Although the data for different sizes exhibit a significant bifurcation as $s$ 
approaches $L$, a clear linear behavior is observed in the intermediate 
region where $s$ satisfies $L \gg s \gg 1$. We performed fits to
the ansatz \eqref{def_Ansatz_ns} by systematically applying different upper 
cutoffs for $s$. For $s \leq 128$, the results are relatively stable, 
yielding $\Omega=0.41(1)$. This result is 1.8 standard deviations from
the conjectured value.

The tricritical one-state Potts model corresponds to  
$g=4/3$, and $\tau=379/187$, and conjecture \eqref{conj_Omega}
yields $\Omega=72/187\approx 0.385\,027$. Since the tricritical one-state
Potts model belongs to the same universality class as the usual Ising model, 
it exhibits a rich symmetry and rather strong finite-size effects. 
Fig.~\ref{fig:6}(c)
shows that the data for different system sizes differ significantly. 
However, after excluding the effects from small and large $s$, a linear 
trend is still observed. Unfortunately, the power-law fit to the ansatz 
\eqref{def_Ansatz_ns} is more challenging. Even if we consider only the data 
for $L=2048$, we could only achieve a stable fit by choosing 
$s_{\rm max}=64$, as reported in Table~\ref{tab:fit2}. We obtain 
$\Omega=0.37(1)$, which is 1.5 standard deviations from the conjectured
value.
	
\section{Discussion}

We studied the cluster-size distribution $n_s$ \eqref{def_ns_c} 
for the two-dimensional O$(n)$ loop model \eqref{def_Z_loop} on the 
branches $x_\pm$ \eqref{def_xplusminus} for $n\in [1,2]$. 
On both branches, the O$(n)$ loop model can be
represented as a Coulomb gas with a certain coupling constant 
$g\in (0,2]$.
Furthermore, on $x_+$ the O$(n)$ loop model is critical, and it belongs to
the same universality class as the tricritical Potts model with $Q=n^2$ 
states. This model has a Coulomb-gas representation with $g\in[1,2]$. In 
addition, the branch $x_-$ is in the dense critical phase of the O$(n)$ 
loop model. This phase belongs to the same universality class as the 
critical Potts model with $Q=n^2$ states. This model has a Coulomb-gas 
representation with $g\in(0,1]$. 
The goal of this article was to gain new insights about the 
correction-to-scaling exponent $\Omega$ in \eqref{def_ns_c}. 

First, we have obtained a closed-form expression for $\Omega$ as a function
of $g$ [cf. \eqref{conj_Omega}] by starting from the 
partition function of the  O$(n)$ model on this annulus 
\cite{Saleur_89,Cardy_06},
and generalizing an argument by Ziff \cite{Ziff_11}. The result is, not 
surprisingly, a rational function of $g$, and indeed it gives the exact 
result for percolation obtained by Ziff.  

We have then confirmed the validity of conjecture \eqref{conj_Omega} by 
performing high-precision Monte Carlo simulations. We have chosen to simulate 
the O$(n)$ loop model rather than the critical or tricritical Potts models. 
There are two main motivations to do so: (1) critical slowing down is absent 
on $x_-$, and is moderate for $Q\in[1,2]$ on $x_+$; and (2) the leading 
exponents are absent in the O$(n)$ model. This implies that the effect of
the higher-order terms not included in the ansatz \eqref{def_Ansatz_ns} 
are smaller than expected.    
We applied a cluster Monte
Carlo algorithm to a generalized Ising model \eqref{def_Z_GIsing} which
is equivalent (modulo boundary conditions) to the O$(n)$ loop model. 

The numerical results fully support conjecture \eqref{conj_Omega} for
$\Omega$ when $Q=1,2,3$ on the $x_-$ branch, $Q=1,2$ on the $x_+$ branch,
and $Q=4$, which belongs to both branches. 
In the description of the fit results (Sec.~\ref{sec:res}), 
it is frequently mentioned that there exists some deviations 
(up to two error bars) between the numerical estimate and the theoretical 
prediction. 
We emphasize that this is a typical phenomenon since potential 
systematic errors are not taken sufficiently into account in the 
analyses, which would in principle need a careful extrapolation for both 
the $s \to \infty$ and the $L \to \infty$ limits.

We believe that the conjecture for $\Omega$ extends to the whole branches
$x_\pm$. This means the interval $Q\in[0,4]$ for both the critical and 
tricritical Potts models, and the interval $n\in[-2,2]$ for the O$(n)$ loop
model. Additional Monte Carlo simulations could be done to see whether 
this conjecture is true. 

It is also worth noticing that the difference between the two most relevant 
magnetic eigenvalues $y_{h1}-y_{h2}$ is related to the exponent $\Omega$ 
(i.e., $y_{h1}-y_{h2}=\Omega\, d_f = \Omega\, y_{h1}$). 
This means that it is possible to obtain the
subleading magnetic exponent by measuring $\Omega$ in e.g., $d$-dimensional
percolation. This observation could be useful to obtain accurate results
for the subleading exponent $y_{h2}$ in other models.  

\begin{acknowledgments}
We thank Robert M. Ziff for fruitful discussions. 
This work has been supported by the National Natural Science Foundation of 
China (under Grant No. 12275263), the Innovation Program for Quantum Science 
and Technology (under grant No. 2021ZD0301900), the Natural Science 
Foundation of Fujian Province of China (under Grant No. 2023J02032). 
\end{acknowledgments}     
	

 \input{main_v3.bbl}
\end{document}

%% file: main_v3.bbl
%